\documentclass[a4paper,11pt]{article}
\usepackage{pos}
\usepackage{lineno}

\title{Underlying-event properties in proton-proton collisions at $\mathbf{\sqrt{s}=13.6}$ TeV measured with the ALICE detector at the LHC}

\ShortTitle{UE properties in pp collisions at $\sqrt{s}=13.6$ TeV}

\author*[a]{Jesús Eduardo Muñoz Méndez}

\onbehalf{for the ALICE Collaboration}

\affiliation[a]{Instituto de Ciencias Nucleares, UNAM\\
 Circuito exterior s/n Ciudad Universitaria, 04510 Mexico City, Mexico}

\emailAdd{jesus.munoz@cern.ch}

\abstract{This contribution reports measurements characterizing the underlying event at midrapidity in pp collisions at $\sqrt{s}=13.6$\,TeV, based on LHC Run 3 data collected by ALICE. The charged-particle number and summed transverse-momentum densities are measured as a function of the leading-particle transverse momentum ($p_{\rm T}^{\rm leading}$) in the toward, transverse, and away regions, defined relative to the azimuthal direction of the leading charged particle. The particle densities in the transverse region are observed to increase steeply at low $p_{\rm T}^{\rm leading}$, followed by a plateau for $p_{\rm T}^{\rm leading}>5$\,GeV/$c$. The results are compared with measurements at lower center-of-mass energies to illustrate the energy dependence of the underlying-event activity, and with predictions from PYTHIA~8 and EPOS4.}

\FullConference{
43rd International Conference on High Energy Physics (ICHEP 2026)\\
30 July to 5 August, 2026\\
Natal, Brazil
}

\begin{document}
\maketitle

\section{Introduction}

In high-energy collisions, final-state particles originate both from hard partonic scatterings and from the underlying event (UE), defined as the collection of particles produced by beam-remnant fragmentation and multiparton interactions (MPI). Disentangling these contributions is essential for properly tuning Monte Carlo event generators~\cite{CMS:2015wcf}, performing precision measurements of hard-process effects~\cite{Sahu:2025mla}, and, more recently, understanding whether collective effects in small systems originate from a deconfined medium whose formation could be driven mainly by MPI. To this end, one can characterize the global features of a proton-proton (pp) collision by studying particle production in different topological regions defined relative to the direction of the particle with the highest transverse momentum in the collision ($p_\text{T}^\text{leading}$), referred to as the leading particle. One can then measure the particle production in the toward, transverse, and away regions, defined relative to the azimuthal direction of the leading charged particle. In general, particles from hard scatterings largely dominate the toward and away regions. By contrast, the transverse region is expected to exhibit greater sensitivity to the underlying event. These three topological regions provide a useful way to measure the bulk event activity and test it against model predictions.

The topological regions are defined by partitioning the azimuthal plane into three regions relative to the leading-particle direction:

\begin{itemize}
\item \textbf{Toward or near region:} $|\Delta\varphi| < 60^\circ$
\item \textbf{Transverse region:} $60^\circ < |\Delta\varphi| < 120^\circ$
\item \textbf{Away region:} $|\Delta\varphi| > 120^\circ$
\end{itemize}

where $\Delta\varphi=\varphi^{\rm leading}-\varphi^{\rm assoc.}$ and "associated" refers to any particle within the defined regions other than the leading particle. Each of these regions can then be characterized by observables such as the number density and the summed-$p_{\mathrm{T}}$ density, defined as

\begin{enumerate}
    \item \textbf{Number density} (average charged-particle density):
    \begin{equation}
    \frac{1}{N_{\text{ev}}(p_{\rm T}^{\text{leading}})}  \frac{1}{\Delta \eta \, \Delta \varphi} N_{\text{ch}}(p_{\rm T}^{\text{leading}}) \, ,
    \end{equation}
    \item \textbf{Summed-$p_{\rm T}$ density} (average summed-$p_{\rm T}$ density):
    \begin{equation}
    \frac{1}{N_{\text{ev}}(p_{\rm T}^{\text{leading}})}  \frac{1}{\Delta \eta \, \Delta \varphi} \sum p_{\rm T}(p_{\rm T}^{\text{leading}}) \, .
    \end{equation}
\end{enumerate}

where $\Delta\phi=2\pi/3$ corresponds to the azimuthal width of the regions, $\Delta\eta=1.6$ is the pseudorapidity acceptance range and $N_{\text{ev}}$ indicates that these observables are averaged over all selected events in each $p_{\rm T}^{\text{leading}}$ interval, including events with zero associated particles in the region. In this study, these observables are based on primary charged particles \footnote{Particles that have a mean proper lifetime greater than 1\,cm$/c$ and which are produced directly in the interaction or from decays of particles with a mean proper lifetime smaller than 1\,cm$/c$~\cite{ALICE:2017hcy}} reconstructed in the pseudorapidity range $|\eta|<0.8$ with $p_{\rm T}^{\rm threshold}>$ 0.15~GeV/$c$. The data set used corresponds to minimum-bias pp collisions at $\sqrt{s}=13.6$ TeV measured with the ALICE detector during Run 3 at the LHC.

\section{Results}

\begin{figure}[tb!]
    \centering
    \includegraphics[width=0.45\textwidth]{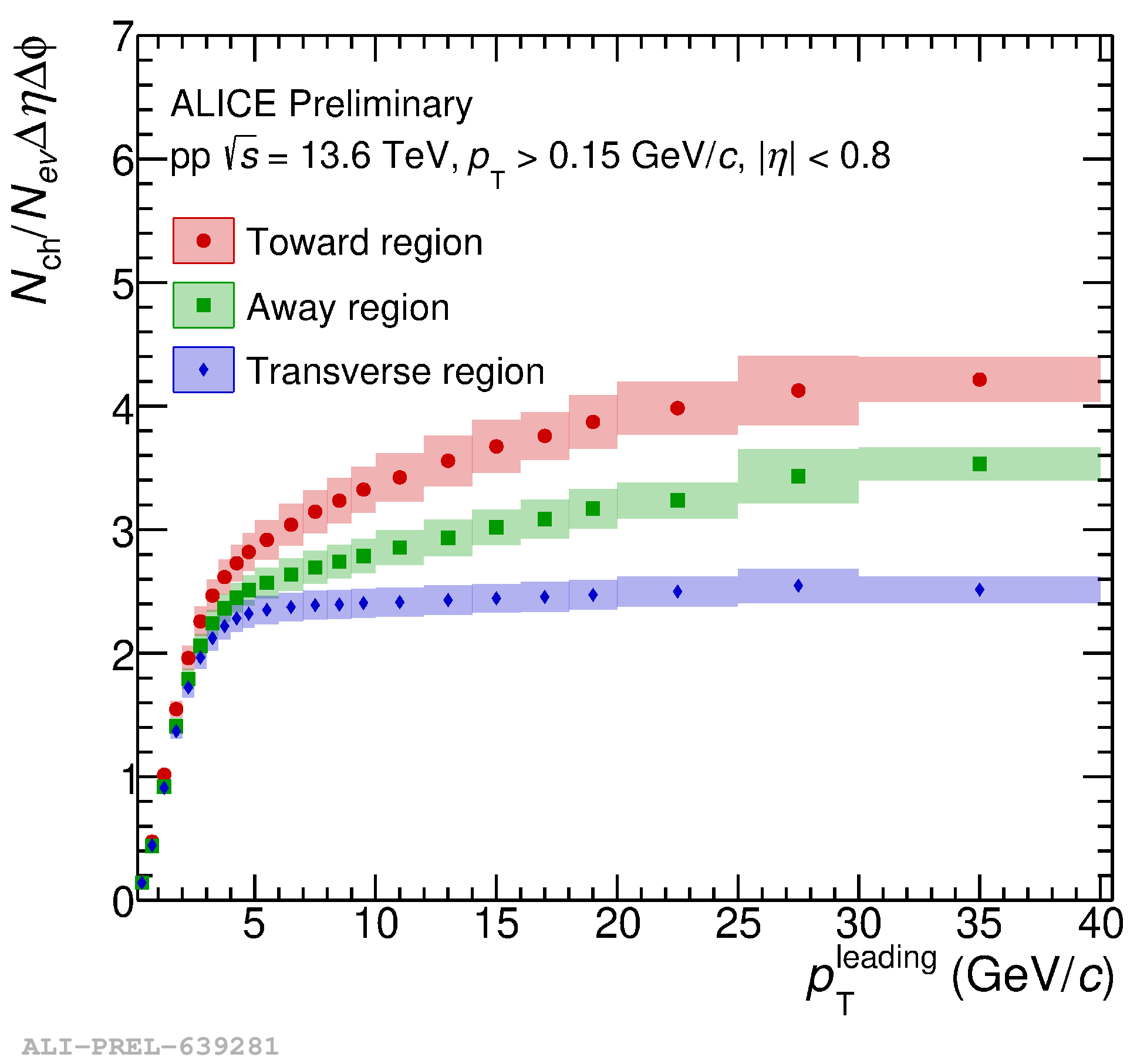}
    \includegraphics[width=0.45\textwidth]{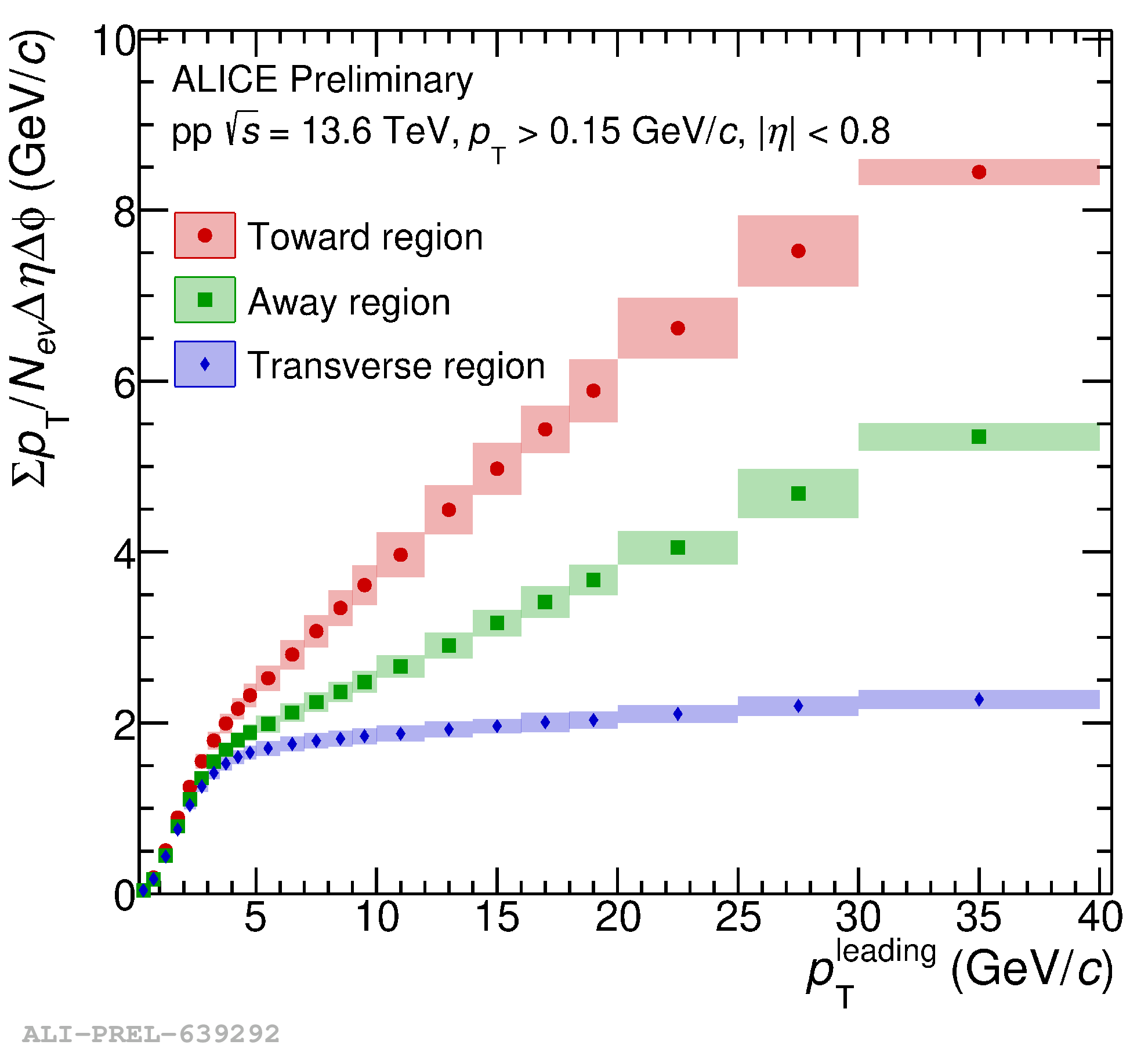}
    \caption{Fully corrected number density (left) and summed-$p_{\rm T}$ density (right) as a function of $p_{\rm T}^{\rm leading}$ for pp collisions at $\sqrt{s}=13.6$ TeV. In each panel, the three topological regions are shown for tracks within $|\eta|<0.8$ and $p_{\rm T}^{\rm track}>0.15$ GeV/$c$. The shaded areas represent the systematic uncertainties, and the vertical error bars indicate the statistical uncertainties.}
    \label{fig:Results}
    \end{figure}

The fully corrected UE observables for tracks with $p_{\rm T}>0.15$ GeV/$c$ are shown in Fig. \ref{fig:Results}. The most prominent feature in both observables is the behavior of the transverse region: the activity increases rapidly at low $p_{\rm T}^{\rm leading}$ and saturates for $p_{\rm T}^{\rm leading}>5$~GeV/$c$. In MPI models, a higher $p_{\rm T}^{\rm leading}$ biases the event sample towards smaller impact parameters and thus towards a larger number of MPIs \cite{Sjostrand:1987su, Strikman:2011cx}. The plateau is consistent with this bias saturating at high $p_{\rm T}^{\rm leading}$, beyond which the selected events have a similar average MPI activity. In contrast, the toward and away regions do not exhibit this saturation, as they receive contributions from both the UE and hard-scattering products, which are predominantly back-to-back in azimuth and whose contribution increases with $p_{\rm T}^{\rm leading}$.

\subsection{Energy dependence}

\begin{figure}[tb!]
    \centering
    \includegraphics[width=\textwidth]{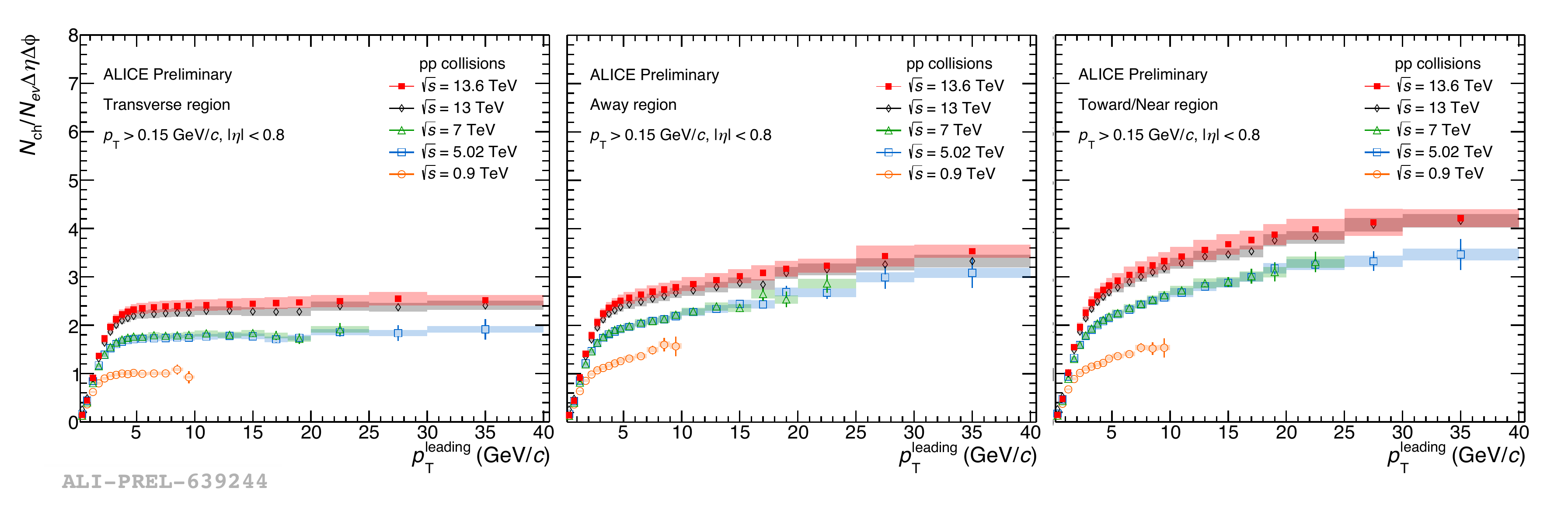}
    \caption{Comparison of the different data sets reported by ALICE \cite{ALICE:2011ac,ALICE:2022fnb,ALICE:2019mmy} for the number density as a function of $p_{\rm T}^{\rm leading}$ in the three topological regions: transverse (left), away (middle), and toward (right). The shaded areas represent the systematic uncertainties, and the vertical error bars indicate the statistical uncertainties.}
    \label{fig:numden_alice}
    \end{figure}

Figure~\ref{fig:numden_alice} compares the number density in the three topological regions with previous ALICE measurements \cite{ALICE:2011ac,ALICE:2022fnb,ALICE:2019mmy}. Across all energies, a common trend is observed: a steep rise followed by a reduced slope, with a plateau reached in the transverse region. Furthermore, a clear center-of-mass-energy dependence is observed, with the activity increasing with $\sqrt{s}$.

\begin{figure}
    \centering
    \includegraphics[width=0.5\textwidth]{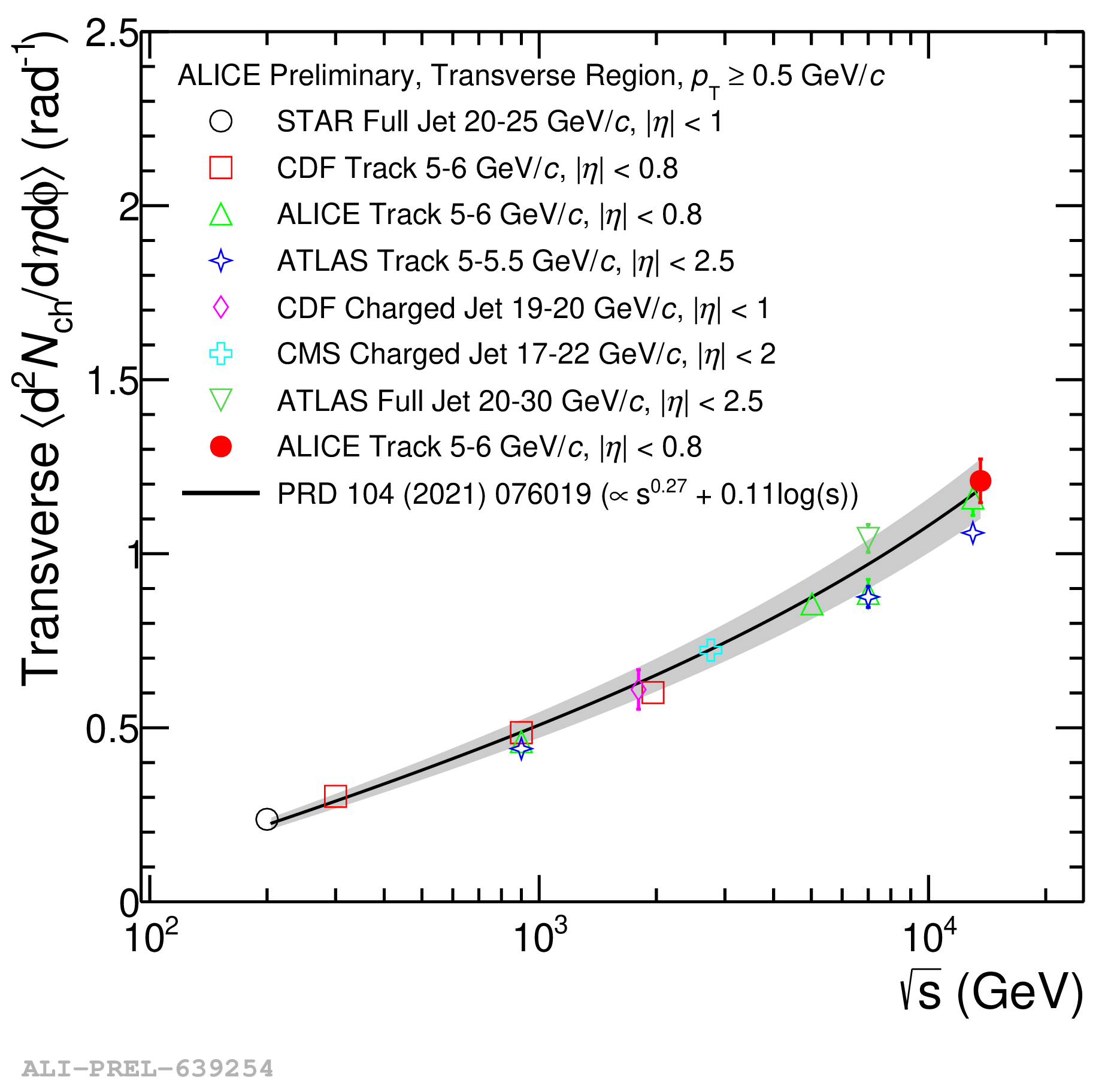}
    \caption{Charged-particle number density in the transverse-region plateau as a function of the center-of-mass energy. Results from different experiments are shown together with a parametrization of the data. Error bars represent statistical and systematic uncertainties summed in quadrature.}
    \label{fig:trans_comp}
    \end{figure}

This dependence on the center-of-mass energy is further illustrated in Fig.~\ref{fig:trans_comp}, which shows the number density in the transverse-region plateau as a function of $\sqrt{s}$. The figure includes results from RHIC \cite{STAR:2019cie}, the Tevatron \cite{CDF:2001onq, CDF:2015txs}, and the LHC \cite{ALICE:2011ac,ALICE:2019mmy,ALICE:2022fnb,ATLAS:2010kmf,ATLAS:2014iez,ATLAS:2017blj,CMS:2015yqj}. For this analysis, the value was calculated in the interval $5<p_{\rm T}^{\rm leading}<6$ GeV/$c$, where the activity is observed to saturate. The ratio between the value reported here and the previous measurement at $\sqrt{s}=13$~TeV, both calculated within the same interval, is $1.043\pm0.070$, which is compatible with unity within uncertainties. The value reported here is also consistent with the parameterization from Ref.~\cite{Ortiz:2021gcr}, obtained considering all the previously reported measurements shown in the figure.

    \begin{figure}[]
    \centering
    \includegraphics[width=\textwidth]{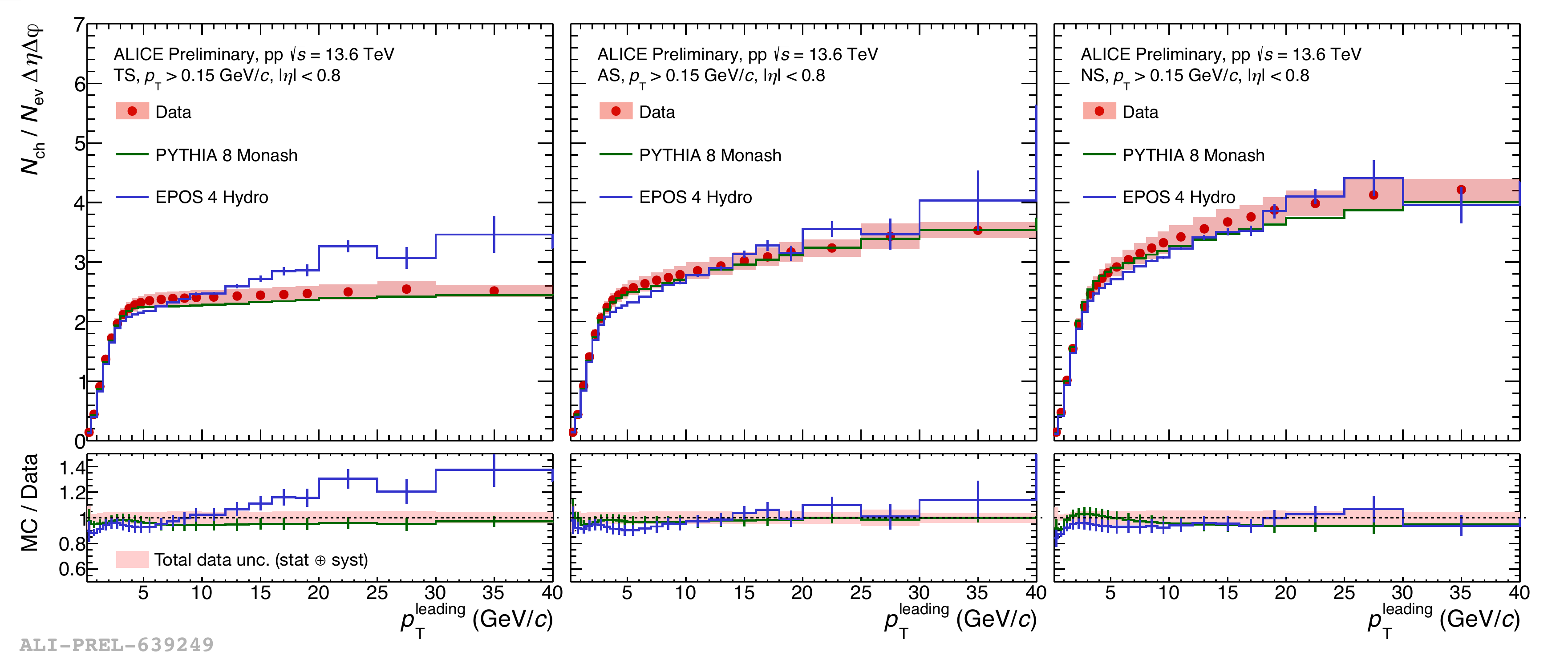}
    \caption{Comparison of PYTHIA~8 and EPOS4 predictions with data for the number density as a function of $p_{\rm T}^{\rm leading}$ in the three topological regions for tracks within $|\eta|<0.8$ and $p_{\rm T}^{\rm track}>0.15$ GeV/$c$. The shaded areas represent the systematic uncertainties, and the vertical error bars indicate the statistical uncertainties.}
    \label{fig:pythia_vs_run3}
    \end{figure}

\subsection{Comparison with MC models}

Figure \ref{fig:pythia_vs_run3} shows a comparison between PYTHIA~8 (tune Monash)~\cite{Bierlich:2022pfr}, EPOS4 with hydrodynamic evolution~\cite{Werner:2024fwk}, and data for the number density as a function of $p_{\rm T}^{\rm leading}$ in the three topological regions. The ratios indicate a good agreement between both event generators and data at low $p_{\rm T}^{\rm leading}$, with deviations up to $\sim10\%$. For higher values of $p_{\rm T}^{\rm leading}$, PYTHIA 8 provides a better description of the UE in all regions, while EPOS4 overestimates the activity in the transverse region.

\section{Summary}

The number density and summed-$p_{\rm T}$ density are reported as a function of $p_{\rm T}^{\rm leading}$ in pp collisions at $\sqrt{s}=13.6$~TeV across three topological regions, defined relative to the direction of the leading particle. While these observables in the transverse region saturate at $p_{\rm T}^{\rm leading}>5$~GeV/$c$, the toward and away regions continue to rise as the contribution from hard-scattering products grows. Comparison with previous ALICE measurements shows a clear dependence on the center-of-mass energy, with the UE activity reported here showing a compatibility within uncertainty with the previously reported measurement at $\sqrt{s}=13$~TeV. Finally, although both PYTHIA~8 and EPOS4 reproduce the data to within $\sim10\%$ at $p_{\rm T}^{\rm leading}<10$~GeV/$c$, at higher $p_{\rm T}^{\rm leading}$ PYTHIA~8 gives a better description of the UE activity in all three regions, while EPOS4 overestimates the transverse-region density. These results indicate the need for further tuning of these event generators to improve their predictions.

\section*{Acknowledgments} 
This work has been supported by DGAPA-UNAM PAPIIT No. IG100524

\end{document}